\documentstyle[twoside,fleqn,espcrc2,epsf]{article}

\newcommand{\AmS}{{\protect\the\textfont2
  A\kern-.1667em\lower.5ex\hbox{M}\kern-.125emS}}

\newcommand{\be}{\begin{equation}}
\newcommand{\ee}{\end{equation}}

\title{Some non-perturbative aspects of gauge fixing in
two-dimensional Yang-Mills theory}
\author{James E. Hetrick%
  \address{Institute for Theoretical Physics\\ University of Amsterdam\\
	Valckenierstraat 65\\ 1108-DX Amsterdam, ~~The Netherlands}}

\begin{document}

\begin{abstract}
Gauge fixing in general is incomplete, such that one solves some of
the gauge constraints (fixes to some gauge), quantizes, then imposes
any residual gauge symmetries (Gribov copies) on the
wavefunctions. While the Fadeev-Popov determinant keeps track of the
local metric on this gauge fixed surface, the global topology of the
reduced configuration space can be different depending on the
treatment of the residual symmetries, which can in turn
affect global properties of the theory such as the vacuum
wavefunction.

Pure $SU(N)$ gauge theory in two dimensions provides a simple yet
non-trivial example where the above structure and effects can be
elucidated explicitly, thus displaying physical effects of
the treatment of Gribov copies.
\end{abstract}
\maketitle

\def\rightmark{\rm \protect\parbox[rt]{2in}{ITFA-93-39 \protect\\
    November 1993}}

\section{Introduction}

The essential nature of a gauge theory is of course its gauge
symmetry, whereby a {\it fiber}\footnote{I will use freely the
language of fiber bundles, which is the natural mathematical setting
of gauge theories. For an excellent introduction to the geometry and
topology of fiber bundles, see \cite{Nakahara}} of
gauge transformed fields desribes a unique physical
configuration. Symbolically we represent this structure with the
following icon
\be
\matrix{{\cal G}&\longrightarrow&{\cal Q}\cr
           ~    &       ~       &~~~\downarrow \Pi\cr
           ~    &       ~       &~~~~~~~~Q_{\rm physical}\cr}
\ee
where the {\it base space} of physical configurations $Q_{\rm
physical}$ would have coordinates $z$, while the fiber bundle ${\cal
Q}$ has coordinates $A_\mu(x)=(z,\Omega)$, with $\Omega$ parametrizing
the fiber. The meaning of ${\cal G}\rightarrow {\cal Q}$ is that the
{\it group} ${\cal G}$, the set of all possible gauge transformations,
acts on the fiber coordinates $(z,\Omega)\rightarrow
(z,\Omega^\prime)$. The {\it projection} $\Pi$ projects from ${\cal
Q}$ down to $Q_{\rm physical}$, $\Pi: (z,\Omega)\rightarrow z$.

Gauge fixing the theory, means finding a unique point up in the bundle
for each physical configuration $z \in Q_{\rm physical}$, ie. an
inverse projection $\Pi^{-1}: z \rightarrow (z,\Omega(z))$, which is
called a {\it section} of the bundle. Classically, variation of the
Lagrangian gives the equations of motion of a trajectory $z(t)$ on the
base space $Q_{\rm physical}$ and any {\it lift} of this trajectory to
a section in the bundle will satisfy the equations of motion. The
choice of this section or gauge, is a matter of computational
convenience for the problem at hand.

In a quantum theory of gauge fields the situation is greatly
changed, since the wave functions, or the measure in the path
integral, are sensitive to the global geometry of the
configuration space, rather than just the local structure of phase space as
in the Hamiltonian evolution of the classical system. This allows for some
ambiguity in the definition of the quantum gauge theory, since
we have the choice of either solving the constraints first,
then quantizing on the reduced configuration space, or quantizing
first, and implimenting the gauge symmetries as constraints (symmetries)
of the wave functions.

Usually we proceed with some combination of the above for technical
reasons, whereby we gauge fix the fields to some convenient gauge, which
leaves residual gauge symmetries still acting on the reduced configuration
space. These we impliment as symmetries on the wave functions. Such are the
residual homotopically non-trivial gauge transformations which generate
the $\theta$ vacua, as well as residual Gribov copies which occur in
non-Abelian gauge theories. Scematically we can represent this structure
with the following diagram:

\be
\matrix{{\cal G}&\longrightarrow&{\cal Q}\cr
           ~    &       ~       &~~~~~~~~~\downarrow
                \Pi_{_{\partial\cdot A = 0}}\cr
 G^\prime \subset {\cal G} &\longrightarrow&~Q^\prime\cr
           ~    &       ~       &~~~~~~~~~\downarrow \Pi_{_{\rm
                Gribov}}\cr
 G^{\prime\prime} \subset G^\prime &\longrightarrow&~~Q^{\prime\prime}\cr
	   ~	&	~	&\vdots\cr
           ~    &       ~       &~~~\downarrow \Pi\cr
           ~    &       ~       &~~~~~~~~Q_{\rm physical}\cr}
\ee

The local geometry (metric) of the successive configuration spaces
${\cal Q}, Q^\prime,Q^{\prime\prime},\dots$ is taken care of by the
Fadeev-Popov determinant as the Jacobian of the projection down to the
reduced configuration space at each stage. This insures that
the perturbative structure is intact for any gauge.

The successive residual symmetries contain detailed information about the
global topology of the fundamental modular domain, $Q_{\rm physical}$,
which is a crucial aspect of the non-perturbative structure of the
theory. As the lattice provides the only non-perturbative means of
calculation at present, it is important to understand the emergence of
the topology of the configuration space, the treatment of the Gribov
copies, and the artifacts in the gauge orbit space due to the lattice.

The quantum theories defined at different levels $Q^\prime,
Q^{\prime\prime},\dots$ are defined on configuration spaces of
different topology, since at each level more of the residual
constraints have been solved and the configuration space compactified
accordingly. These different topologies of $Q^\prime$ and
$Q^{\prime\prime}$ will affect the types of Hilbert spaces which can
be constructed over them. For instance, a general wave function on a
circle $S^1$ can have a phase $e^{i\alpha}$ upon transport around the
circle. Although this circle is a submanifold of an $n-$sphere, and
hence obtained by some projection $\Pi_{S^n\rightarrow S^1}$, wave
functions on $S^n$ must be single valued since $\pi_1(S^n) = 0$ for $n
> 1$.  In general we may find that the topologies of $Q^{\prime}$ and
$Q^{\prime\prime}$ dictate incomensurate Hilbert spaces, in which case
we have inequivalent quantizations of the gauge theory.  Such is the
case for $SU(N)$ gauge fields in two spacetime dimensions
\cite{R,Joakim,CE}.

\section{$SU(N)$ gauge theory in two dimensions}

We consider periodic $SU(N)$ gauge fields $A_\mu(x)$ on a cylindrical
spacetime $(x,t) \in S^1\times R^1$ using canonical quantization to
illuminate the different Hilbert spaces that emerge depending on the
level of gauge fixing; details can be found in \cite{R}.

First we identify the fundamental modular domain $Q_{\rm physical}$
explicitly, which is possible due to the simple structure of a gauge
theory in two dimensions. As explained in \cite{R}, the set of
physically distinct field configurations is isomorphic to the space of
conjugacy classes of $G=SU(N)$. This space is an orbifold, which we
label $\Xi_G$, made by identifying the points in the maximal
torus of $G$ (the Cartan subgroup) under reflections in the
diagonal planes of the torus: $\theta_i \leftrightarrow
\theta_j$. Since these reflections invert the tangent plane,
$\Xi_G$ will be {\it non-orientable}, ie. it has no globally
defined tangent bundle. This is an essential feature of the global
topology of the fundamental modular domain and will affect the types
of wave functions it admits.

For a concrete example consider $G=SU(2)$. In fact we can almost fix
the gauge completely to $Q_{\rm physical}$ if $A_1$ satisfies
the Coulomb gauge condition, $\partial_1 A_1 = 0$, and is further
diagonal. Then
\be
A_1(x,t) = \Big(\matrix{\theta(t)&~\cr
	 	   ~& -\theta(t)\cr}\Big)
\ee
corresponding to a configuration space $Q^{\prime\prime}$ just above
the fundamental moduli space $Q_{\rm physical}$
with only discrete residual symmetries left,
namely: $\theta \rightarrow \theta+n\pi$ and $\theta \rightarrow
-\theta$. It is the latter transformation which makes the circle
defined by $\theta$ into the non-orientable orbifold $\Xi=(0,\pi)_\pm$.
By $(0,\pi)_\pm$ is meant that $\Xi_{SU(2)}$ is described by two
patches $(0,\pi)$ which cover the same points but have oppositely
oriented tangent spaces.

In this gauge we easily find that the Hamiltonian is simply the
Laplacian on the circle
\be
H = -\frac{g^2L}{8}\frac{\partial^2}{\partial \theta^2}
\ee
where $g$ is the coupling constant and $L$ is the volume of space.  In
general the wave function $\Psi_n(\theta)$ could be some combination
of $\exp[\pm i2\pi\alpha(\theta + \beta)]$, however non-integer
$\alpha$ implies some some physics (Bohm-Aharonov phenomena) at
$\theta =0$ or $\pi$. That $\Xi_{SU(2)}$ is non-orientable in fact
determines $\Psi_n \sim \cos(n\theta)$ since it is this wave function
whose momentum vanishes at the orbifold singularities where the
tangent space is ill defined \cite{FG,Dowker}. The spectrum in this
case is
\be
E_j = j^2
\ee
and $j=0,\frac{1}{2},1,\frac{3}{2},\dots$

If on the other hand we gauge fix only to the Coulomb gauge condition,
$\partial_1 A_1 = 0$, the configuration space is isomorphic to the
group manifold $SU(2) = S^3$ itself, and the residual gauge symmetries
are all constant adjoint rotations of $A_1$. Then the Hamiltonian is
the Laplacian on $SU(2)$ and wave functions are characters
$\chi_j(\theta)$, which are invariant under adjoint action of $G$ on
itself. In comparing the spectrum on $G$ to the one above (on the flat
maximal torus), there is a constant shift in energy proportional to
the curvature of $G$. This then yields the spectrum
\be
E_j = j(j+1) + \frac{1}{4} = (j+\frac{1}{2})^2
\ee
with $j$ taking half integer values as above. The two spectra agree
except for the vacuum states as depicted in figure 1.
\begin{figure}[htb]
\epsfxsize= 7.3cm
\epsffile[190 375 470 540]{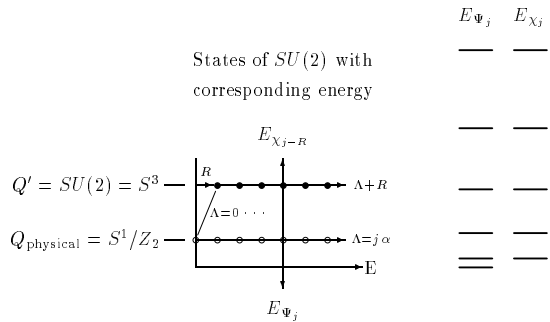}
\caption{Correspondence of the excited spectra and the shift of
ground state for two different gauge choices. States are labeled by
irreducible representations $\Lambda = j\cdot\alpha$ of $SU(2)$, where
$\alpha$ is the normalized root vector. $R$ is the shift
due to the curvature of $G$ and equals half the sum of the positive
roots. See [2] for further details.}
\label{fig:1}
\end{figure}

The difference is that quantizing on the maximal
torus, preserving the non-orientable nature of $Q_{\rm physical}$
chooses even (cos) fourier modes on the maximal torus, However only
the odd (sin) modes have a globally defined lift to the gauge group
manifold. Characters
\be
\chi_j = \frac{\sin[(j+1/2)\theta]}{\sin(\theta/2)}
\ee
are odd fourier modes, $sin[(j+1/2)\theta]$, divided by the square
root of the Haar measure (also odd). This construction is general and
the shift of $j$ by $1/2$ is so that $\chi_{j=0} = 1$. The lift of the
even (cos) mode: $cos(j\theta)/sin(\theta/2)$ is in fact an
eigenfunction of the Laplacian with eigenvalue $j(j+1)$, and while
square integrable, is not a single valued function on $SU(2)$.

The lattice, essentially due to it's compact formulation on $G$ from
the start, gives the latter quantization above, without gauge fixing.
It will be very interesting to see what type of gauge fixing is
necessary in the lattice theory to reproduce the former spectrum, and
if this can indeed be done in the compact theory. Such a program is
underway.

This work is supported by the Stichting voor Fundamenteel Onderzoek
(FOM).

\def\rightmark{}

\end{document}